# Optical net gain measurement on Al$_{0.07}$Ga$_{0.93}$N/GaN multi-quantum well

**QUANG MINH THAI,**[1,*] **SERGI CUESTA,**[2] **LOU DENAIX,**[2] **SYLVAIN HERMELIN,**[1] **OLIVIER BOISRON,**[1] **STEPHEN T. PURCELL,**[1] **LE SI DANG,**[3] AND **EVA MONROY**[2]

[1] *Institut Lumière Matière, CNRS, University of Lyon, Univ. Claude Bernard Lyon 1, F-69622 Villeurbanne, France*
[2] *Univ. Grenoble-Alpes, CEA, Grenoble INP, IRIG, PHELIQS, 17-avenue des Martyrs, F-38000 Grenoble, France*
[3] *Univ. Grenoble-Alpes, CNRS, Institut Néel, 25-avenue des Martyrs, F-38000 Grenoble, France*
*\* [thaiquangminh1993@gmail.com](mailto:thaiquangminh1993@gmail.com)*

**Abstract:** We present net gain measurements at room temperature in Al$_{0.07}$Ga$_{0.93}$N/GaN 10-period multi-quantum well emitting at 367 nm, using the variable stripe length method. Measurements were conducted at two different positions on the sample, where the net gain threshold was attained at 218 kW.cm$^{-2}$ and 403 kW.cm$^{-2}$. At the position with higher threshold, we observed an anomalous amplification of the photoluminescence intensity that occurs for long stripe lengths (superior to 400 µm) and high pumping power (superior to 550 kW.cm$^{-2}$), leading to an overestimation of the gain value. We attribute such a phenomenon to the feedback provided by the reflection from cracks, which were created during the epitaxial growth due to the strong lattice mismatch between different layers. The highest gain value without anomalous amplification was 131 cm$^{-1}$, obtained at the maximum pumping power density of the experimental setup (743 kW.cm$^{-2}$). Using the intrinsic efficiency limit of the cathodoluminescence process, we estimate a lower limit for electron beam pumped laser threshold at room temperature of 390 kW.cm$^{-2}$ for this multi-quantum well structure.

## 1. Introduction

Ultraviolet (UV) lasers find application in many domains, like germicidal disinfection and medical treatments, as well as in Lidar remote detection and non-line-of-sight (NLOS) communication. The demand on UV lasers is currently covered by excimer lasers (ArF, Ar$_2$, XeCl, ...) and lasers based on frequency conversion (Nd:YAG). These lasers come with several regulations and specific safety instructions: For example, toxic gases are involved in the operation of excimer lasers, in both the injection phase (F$_2$, Xe, ....) and the lasing phase, with the release of O$_3$ as the by-product of the gas ionization process. Frequency-conversion lasers rely on non-linear processes to operate in UV wavelength range, which often have to deal with stability issues and the ageing of optics, with some being highly hygroscopic. Wide band-gap semiconductors, like AlGaN alloys, could provide an alternative for electrical-injected UV lasers, with tunable wavelengths which depend on the Al molar fraction. However, the requirement of low resistivity and high hole concentration for p-type AlGaN poses a challenge for the development of such devices, especially in the case of laser diodes, where high injected current densities of tens of kA.cm$^{-2}$ are required to achieve stimulated emission [1]–[8].

Recently, electron beam pumping was proposed as an alternative solution to inject electrons and holes into the AlGaN based active region, obviating the need for p-type AlGaN. In this approach, high-energy electrons are first accelerated from the cathode to the sample, and then transmit their kinetic energy to valence band electrons via impact-ionization processes, generating electron-hole pairs. Experimental demonstrations of UV lasing with this method were reported by Hayashi *et al.* for an $Al_{0.07}Ga_{0.93}N$/GaN multi-quantum well (MQW) at 107 K and 230 kW.cm$^{-2}$ lasing threshold [9], and by Wunderer *et al.* for InGaN quantum wells within a GaN/AlGaN heterostructure at 77 K and 159 kW.cm$^{-2}$ [10]. Room temperature lasing has not yet been observed, however. It is therefore interesting to conduct other characterization at room temperature on this kind of structures, like a quantification of the net gain, to better understand its optical properties and have an idea of the expected electron beam pumped laser threshold, if it occurs.

The paper is organized as follows: in section 2, we present the structure of the $Al_{0.07}Ga_{0.93}N$/GaN MQW, with its epitaxial growth and structural characterization. The sample resembles closely to the one described in Ref. [9] for the demonstration of an electron-beam pumped laser. The optical setup for net gain measurements, using the variable stripe length (VSL) method [11], will also be described in this section. In section 3, we show the PL spectra and the net gain calculation at two different positions of the sample, revealing significant different results. In section 4, we discuss the origin of such a difference, and finally, we estimate a lower limit for electron beam pumped laser threshold at room temperature for this structure, by combining the net gain threshold with the theoretical limit of electron-hole generation efficiency in the cathodoluminescence process.

## 2. Sample and experimental setup

The sample under study consists of a 10-period $Al_{0.07}Ga_{0.93}N$/GaN MQW, sandwiched between two layers of $Al_{0.07}Ga_{0.93}N$ as top/bottom inner claddings, and two layers of $Al_{0.15}Ga_{0.85}N$ as top/bottom outer claddings, all grown on top of a bulk GaN substrate using plasma-assisted molecular beam epitaxy. A schematic of the sample is presented in **Fig. 1(a)**. The detailed growth condition can be found in Ref. [12]. The layer thicknesses of the $Al_{0.07}Ga_{0.93}N$/GaN MQW (11.4 nm $Al_{0.07}Ga_{0.93}N$/ 2.8 nm GaN) are very similar to those used by Hayashi et al. (nominally 12 nm $Al_{0.07}Ga_{0.93}N$ /3 nm GaN) for the demonstration of UV lasing under electron beam pumping [9]. However, the waveguide was modified, aiming to increase the optical confinement factor in the MQW and reduce the required accelerating voltage [13]. We depict, in **Fig. 1(b)**, a simulation of the optical field distribution in the first optical mode confined in the heterostructure, obtained from a commercial finite-element analysis software (Comsol Multiphysics). The values of refractive indices were taken from Refs [14], [15]. From such calculation, the optical confinement factor is estimated about 7%. The periodicity of the MQW was analyzed by a high-resolution X-ray diffraction (HRXRD) in a Rigaku SmartLab diffractometer. A $\theta - 2\theta$ X-ray scan around the (0002) reflection of GaN is presented in **Fig. 1(c)**, together with a theoretical calculation based on the structure shown in **Fig. 1(a).**

We conducted net gain measurements at room temperature, using the VSL method described in Ref. [11], with the experimental setup illustrated in **Fig. 2**. The pump source was a 193 nm ArF excimer laser (Compex 102 from Coherent), operated in pulse mode (20 ns pulse width, with a frequency capped at 20 Hz). Such pump laser was already used in several studies of optical gain in AlGaN-based MQW [16]–[18]. At this wavelength, using the absorption coefficients reported by Muth *et al*. [19], 90% of the pumping power is absorbed in the topmost 77 nm of the sample. Photogenerated carriers have to thermalize and diffuse to the active MQW to contribute to the emission. Since the ArF laser beam dimension is relatively large (10 mm × 24 mm), to create a narrow laser stripe on the sample, we insert a rectangular slit (0.05 mm × 3 mm) in front of the laser beam, then recreate its image on the sample using two fused silica spherical lens. The laser stripe on the sample has a width of 25 μm and a maximum length of 1500 μm. A mobile shutter is positioned in front of the sample to modify the stripe length. Both the sample and the laser stripe can be imaged by reflection on a CMOS camera (Thorlabs

DCC1240M). The beam intensity can be modified using fused silica reflective optical densities. PL signals are collected using an optical fiber and then analyzed with a Horiba Triax 320 spectrometer (grating of 1200 grooves/mm, with blaze wavelength at 330 nm), coupled with a cooled CCD camera (Symphony STE).

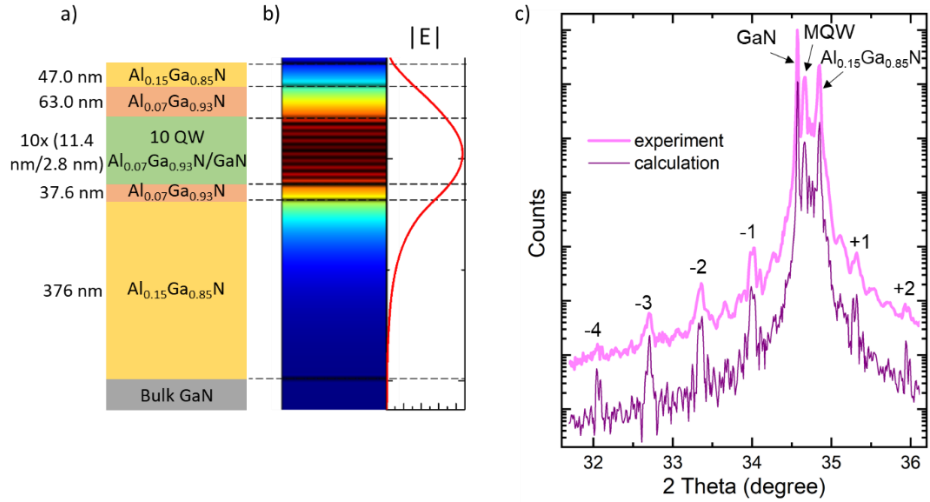

Fig. 1. a) Schema of the $Al_{0.07}Ga_{0.93}N$/GaN MQW structure under study. b) Simulation of the optical field distribution in the first optical mode. c) HRXRD $\theta - 2\theta$ scan around the (0002) reflection of GaN, together with a theoretical calculation for the structure shown in (a), assuming 50% relaxation of the $Al_{0.15}Ga_{0.85}N$ bottom outer cladding layer. Labels indicate the original diffraction peaks of the bulk GaN substrate, the $Al_{0.15}Ga_{0.85}N$ layers and the $Al_{0.07}Ga_{0.93}N$/GaN MQW, along with the order of its satellite peaks.

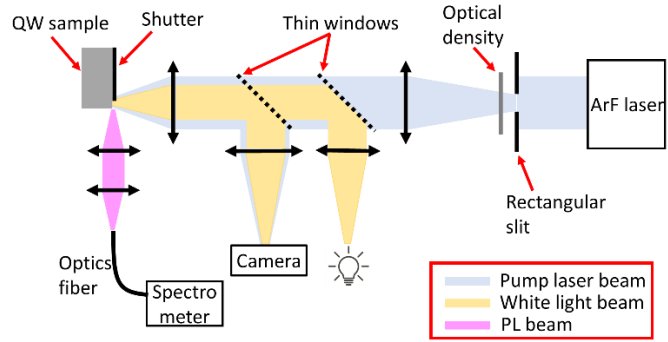

Fig. 2. Experimental setup for net gain measurements using the VSL method.

## 3. Results

**Figures 3(a) and (b)** describe the PL spectra recorded at positions 1 and 2, respectively, measured with a pumping power of 743 kW.cm$^{-2}$, as a function of the stripe length between 40 µm and 400 µm. At position 1 (**Fig. 3(a)**), the PL intensity increases systematically as the stripe length increases, with the highest signal observed for a length of 400 µm. At position 2 (**Fig. 3(b)**), the behavior is similar for stripes shorter than 240 µm. Then, the PL intensity increases rapidly in the range of 240-360 µm, and the shape and width of the PL peak changes. Finally, for the 400 µm long stripe, the intensity saturates. The evolution of the full width at half maximum (FWHM) of the PL peak is shown in **Fig. 3(c).** For stripes longer than 240 µm, the FWHM at position 2 becomes narrower than that at position 1, decreasing down to 1.5 nm at position 2 compared to 2.6 nm at position 1. The superlinear increase of the PL intensity as a function of the stripe length, together with the spectral narrowing, confirm the presence of amplified spontaneous emission (ASE).

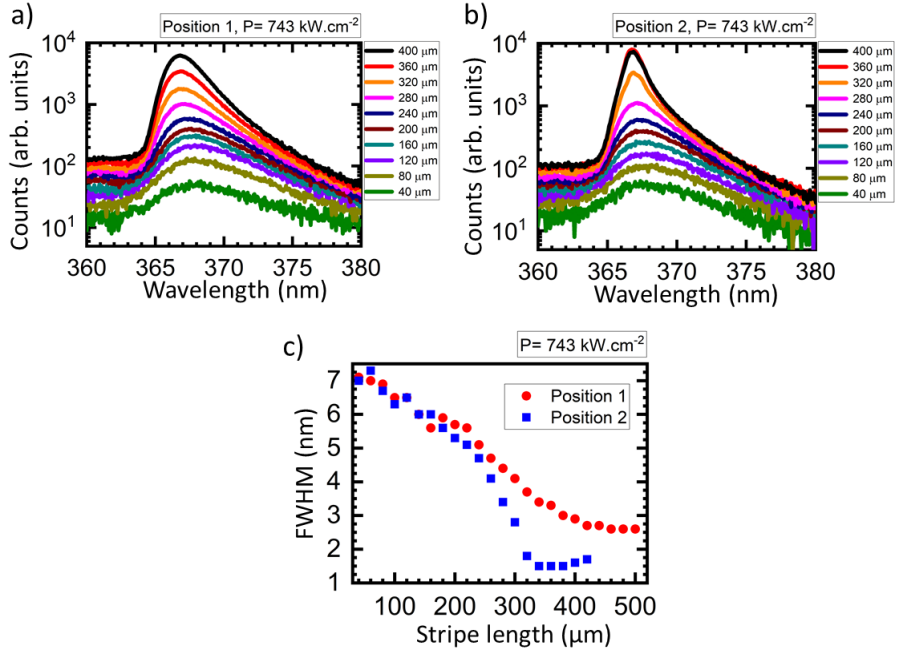

Fig. 3. PL spectra obtained from the $Al_{0.07}Ga_{0.93}N/GaN$ MQW structure for different stripe lengths, at (a) position 1 and (b) position 2. The stripe lengths are indicated in the legends. (c) PL peak FWHM at position 1 and position 2, as a function of the stripe length.

In the VSL method [11], the variation of the PL intensity $I_\lambda$ at a certain wavelength $\lambda$, as a function of the stripe length $l$, is described by a one-dimensional optical amplifier model:

$$I_\lambda(l) = \frac{I_s A}{g_\lambda}(e^{g_\lambda l} - 1) \quad (1)$$

where $I_s$, $A$, $g_\lambda$ are the spontaneous emission rate per unit volume, the cross-section area of the excited volume and the net gain at $\lambda$, respectively. By plotting $I_\lambda(l)$ and then fitting it with the equation (1), we can extract the optical net gain $g_\lambda$.

**Figure 4(a)** shows, for the position 1, the evolution of the PL intensity at $\lambda = 367$ nm as a function of the stripe length, for pumping powers between 218 kW.cm$^{-2}$ and 743 kW.cm$^{-2}$ (indicated in the figure panels). The fitted net gain is –24 cm$^{-1}$ (absorption) at 218 kW.cm$^{-2}$, which increases to 129 cm$^{-1}$ at 743 kW.cm$^{-2}$. The curve shape is different between the absorption cases, where the signal starts to saturate at long stripe length, and in the case of net gain, with an exponential-like increase of the PL intensity. By repeating the fitting process at different wavelengths, we plot, in **Fig. 4(b)**, the net gain as a function of wavelength and pumping power. Positive net gain exists for pumping power higher than 218 kW.cm$^{-2}$, with the highest net gain equal to 131 cm$^{-1}$ ($\lambda = 366.5$ nm) for a pumping power of 743 kW.cm$^{-2}$. At 218 kW.cm$^{-2}$, net gain still exists, but with very low values (3.1 cm$^{-1}$ at $\lambda = 361$ nm). The peak net gain blue shifts from 371 nm at 218 kW.cm$^{-2}$ to 366.5 nm at 743 kW.cm$^{-2}$, as expected due to band filling. The response observed for wavelengths shorter than 365 nm might stem from the mixed optical response of the top claddings (47 nm thick $Al_{0.15}Ga_{0.85}N$ layer and 63 nm thick $Al_{0.07}Ga_{0.93}N$ layer) and the MQW.

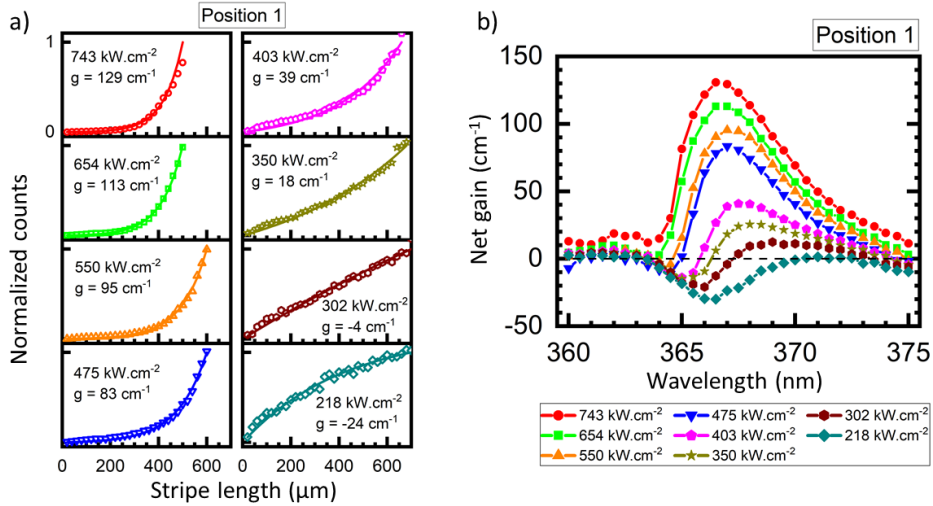

Fig. 4. Net gain measurements performed at position 1: a) PL intensity at λ= 367 nm, as a function of the stripe length, for pumping power between 218 kW.cm$^{-2}$ and 743 kW.cm$^{-2}$. Raw data are presented as symbols and fits are presented as solid lines. The pumping power and the gain value extracted from the fit are shown in each corresponded graph. b) Net gain as a function of wavelength, for pumping powers between 218 kW.cm$^{-2}$ and 743 kW.cm$^{-2}$. Positive values indicate the presence of net gain, while negative values indicate absorption.

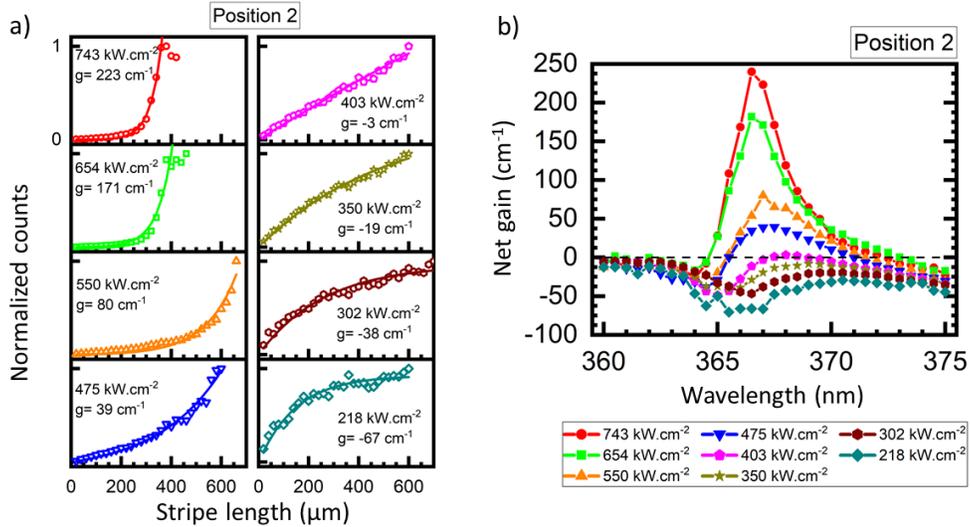

Fig. 5. Net gain measurements performed at position 2: a) PL intensity at λ= 367 nm, as a function of stripe length, for pumping power between 218 kW.cm$^{-2}$ and 743 kW.cm$^{-2}$. Raw data are presented as symbols and fits are presented as solid lines. b) Net gain as a function of wavelength, for pumping power between 218 kW.cm$^{-2}$ and 743 kW.cm$^{-2}$.

We repeated the same net gain measurement at position 2. Compared to position 1, here, at high pumping powers, the PL intensity at 367 nm shows a sharper increase followed by a saturation at high values of stripe length (**Fig. 5(a)**, measurements at 654 kW.cm$^{-2}$ and 743

kW.cm$^{-2}$). The saturation part of the graphs was excluded (if present) during the fit process to extract the gain. The net gain curves as a function of wavelength and pumping power are plotted in **Fig. 5(b)**. In this position, the net gain threshold is 403 kW.cm$^{-2}$, higher than that at position 1 (218 kW.cm$^{-2}$). However, the maximum net gain at 743 kW.cm$^{-2}$ is 240 cm$^{-1}$, significantly higher than that at position 1 (131 cm$^{-1}$). The shape of the net gain curve changes for the three highest pumping powers (550 kW.cm$^{-2}$, 654 kW.cm$^{-2}$ and 743 kW.cm$^{-2}$), resembling a peak rather than the rounded profile expected in net gain measurements. The net gain maximum as a function of the pumping power is plotted in **Fig. 6**, for positions 1 and 2.

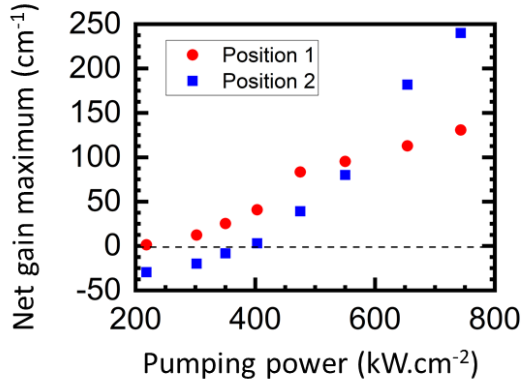

Fig. 6. Net gain maximum as a function of the pumping power, plotted for positions 1 and 2.

## 4. Discussion

The validity of equation (1) relies on several assumptions. First, we assume a constant gain coefficient along the stripe. This condition is not verified in the case of gain saturation, which occurs when the ASE intensity induces a depletion of carriers in the excited state, as the PL intensity builds up along the stripe [20]. Such a phenomenon can create a saturation of PL intensity at long stripes, as observed above at position 2 (**Fig. 5(a)**), which must therefore be excluded in the gain fitting process with equation (1). In addition, the crystal quality itself can be inhomogeneous along the laser stripe. A configuration where the longer stripes reach an area with lower crystal quality, i.e. lower optical gain, could also reproduce the saturation effect.

Second, as discussed in Ref. [20], at low pumping power, the weaker gain gradient and pump-induced refractive index gradient can generate a larger optical mode spreading into the unexcited area of the sample. In this case, inhomogeneous coupling of PL signal can take place, affecting the light collection efficiency and introducing error in the calculation of gain, which is particularly critical near the net gain threshold (here, around 218 kW.cm$^{-2}$). However, it is worth noting that all pumping powers in this study remain in the range of hundreds of kW.cm$^{-2}$, i.e. in the regime of strong optical excitation, and thus maintain a sufficiently strong guiding effect to validate the one-dimensional optical amplifier model, described by the equation (1). Therefore, the error in gain values related to this effect should be minimal.

As described previously in section 3, there are distinct optical responses at the position 2, with significant higher maximum net gain at 743 kW.cm$^{-2}$ (240 cm$^{-1}$), narrower PL spectra, and the presence of gain saturation at long stripe lengths, compared to position 1. To better understand the mechanisms behind such a difference, we analyzed the evolution of the emission characteristics (intensity and FWHM) as a function of the pumping power for a fixed stripe length of 400 μm, for both positions. Results are presented in **Fig. 7(a)-(b)**. The behavior is similar for pumping power lower than 550 kW.cm$^{-2}$. At higher pumping power, for position 2, there is a large enhancement of the PL intensity and a collapse of the FWHM, which together suggest the onset of lasing that would be expected for a material with gain, within an optical cavity. Such a phenomenon is not expected in a sample with only one cleaved facet. However, the image of our sample under an optical microscope reveals the presence of cracks that propagate along the *m*-planes (see **Fig. 7(c)**). These cracks are generated in AlGaN layers grown under tensile stress, as it is the case here due to the lattice mismatch with the GaN substrate. Partial reflection from such extended defects could provide the additional feedback that amplifies further the PL intensity, resulting in the anomalous behavior of the emission characteristics observed at position 2. In this case, care must be taken for the interpretation of net gain using the equation (1), since the effective stripe length might increase due to multiple round trips of photons between cracks, or between a crack and a cleaved edge.

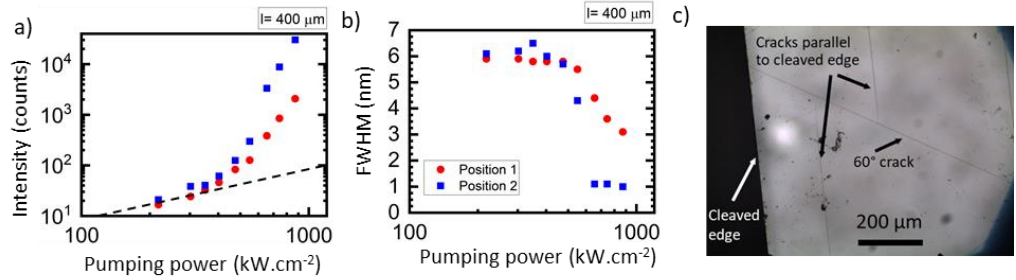

Fig. 7. Variation of (a) the PL intensity and (b) the PL peak FWHM as a function of the pumping power measured at position 1 and position 2, for a stripe length of 400 µm. (c) Top view of the sample in the proximity of the cleaved edge. Cracks propagate either parallel to the cleaved edge or forming a 60° angle.

In the case of an optical cavity with very low mirror loss (for example in very long Fabry-Perot cavities), the presence of optical net gain can be used as an indicator to estimate the threshold pumping power required for lasing. The theoretical limit of the electron-hole generation efficiency in the cathodoluminescence process is approximately 33%, i.e. the kinetic energy of the pump electron should equal at least three times the band gap energy to generate one electron-hole pair [21]. On the other hand, the energy of the pump laser used in the above-described experiments (6.42 eV) is significantly higher than the band gap of the absorbing material (approximately 3.8 eV), which limits the optical pumping efficiency to 59%. Combining these data with the lowest net gain threshold (218 kW.cm$^{-2}$) observed in this work, we estimate, for this structure, a lower limit of $P_{lim} = \frac{0.59}{0.33} 218 = 390$ kW.cm$^{-2}$ for the electron beam pumped laser threshold at room temperature. This predicted threshold is in the range of the current state of the art of UV laser diodes (see **Table 1**). For a real device, specific values of mirror loss can be projected into the mapping of net gain maximum, as a function of the optical pumping power (see **Fig. 6**), to predict the corresponding electron beam pumped laser threshold.

Table 1. Reported values of lasing wavelength λ, threshold current $I_{th}$ and voltage at threshold $V_{th}$ for several UV laser diodes. The lasing threshold power $P_{th}$ (in kW.cm$^{-2}$) was calculated as $P_{th} = I_{th} V_{th}$.

| Reference | λ (nm) | $I_{th}$ (kA.cm$^{-2}$) | $V_{th}$ (V) | $P_{th}$ (kW.cm$^{-2}$) |
|---|---|---|---|---|
| Yoshida *et al.* [1] | 359.6 | 8.0 | 18.0 | 144.0 |
| Yoshida *et al.* [2] | 342.0 | 8.7 | 25.0 | 217.5 |
| Sakai *et al.* [3] | 278.9 | 19.6 | 14.5 | 284.2 |
| Zhang *et al.* [4] | 271.8 | 25.0 | 13.8 | 345.0 |
| Yoshida *et al.* [5] | 336.0 | 17.6 | 34.0 | 598.4 |
| Sato *et al.* [6] | 298.0 | 25.0 | 34.4 | 860.0 |
| Aoki *et al.* [7] | 356.6 | 56.0 | 20.0 | 1120.0 |
| Sato *et al.* [8] | 298.0 | 41.0 | 27.8 | 1139.8 |
|  |  | 67.0 | 27.8 | 1862.6 |

## 5. Conclusion

We reported on optical net gain measurements at room temperature in Al$_{0.07}$Ga$_{0.93}$N/GaN multi-quantum well emitting at 367 nm, using variable stripe length method. The net gain threshold varies between 218 kW.cm$^{-2}$ and 403 kW.cm$^{-2}$, depending on the sample probed position. We observed an anomalous amplification that occurs for long stripe lengths (superior to 400 µm) and high pumping powers (superior to 550 kW.cm$^{-2}$) at the position with higher net gain threshold. We attributed this phenomenon to the optical feedback provided by reflection at cracks, which appear due to plastic relaxation of misfit strain. The highest net gain value without anomalous amplification was 131 cm$^{-1}$, obtained at a pumping power of 743 kW.cm$^{-2}$. By combining the net gain threshold with the theoretical limit of electron-hole generation efficiency in the cathodoluminescence process, we predict, for this structure, a lower limit of 390 kW.cm$^{-2}$ for the electron beam pumped laser threshold at room temperature, which is in the range of current reported values on ultraviolet laser diodes.

**Funding**

This work is supported by the French National Research Agency (ANR) via the UVLASE project (ANR-18-CE24-0014), and by the Auvergne-Rhône-Alpes region (PEAPLE grant).

**Acknowledgements**

The authors would like to thank the platform PLYRA of Institut Lumière Matière - University Claude Bernard Lyon 1 for their support of scientific equipment during this work.

**Disclosures**

The authors declare no conflicts of interest.

**Data availability**

Data underlying the results presented in this paper are not publicly available at this time but may be obtained from the authors upon reasonable request.

# References

1. H. Yoshida, M. Kuwabara, Y. Yamashita, Y. Takagi, K. Uchiyama, and H. Kan, "AlGaN-based laser diodes for the short-wavelength ultraviolet region," *New J. Phys.*, vol. 11, no. 125013, 2009.
2. H. Yoshida, Y. Yamashita, M. Kuwabara, and H. Kan, "A 342-nm ultraviolet AlGaN multiple-quantum-well laser diode," *Nat. Photonics*, vol. 2, pp. 551–554, 2008.
3. T. Sakai, M. Kushimoto, Z. Zhang, N. Sugiyama, L. J. Schowalter, Y. Honda, C. Sasaoka, and H. Amano, "On-wafer fabrication of etched-mirror UV-Claser diodes with the ALD-deposited DBR," *Appl. Phys. Lett.*, vol. 116, no. 122101, 2020.
4. Z. Zhang, M. Kushimoto, T. Sakai, N. Sugiyama, L. J. Schowalter, C. Sasaoka, and H. Amano, "A 271.8 nm deep-ultraviolet laser diode for room temperature operation," *Appl. Phys. Express*, vol. 12, no. 124003, 2019.
5. H. Yoshida, Y. Yamashita, M. Kuwabara, and H. Kan, "Demonstration of an ultraviolet 336 nmAlGaN multiple-quantum-well laser diode," *Appl. Phys. Lett.*, vol. 93, no. 241106, 2008.
6. K. Sato, K. Yamada, S. Ishizuka, S. Yasue, S. Tanaka, T. Omori, S. Teramura, Y. Ogino, S. Iwayama, H. Miyake, M. Iwaya, T. Takeuchi, S. Kamiyama, and I. Akasaki, "AlGaN-based ultraviolet-B laser diode at 298 nm with threshold current density of 25 kA/cm2," *2020 IEEE Photonics Conf.*, 2020.
7. Y. Aoki, M. Kuwabara, Y. Yamashita, Y. Takagi, A. Sugiyama, and H. Yoshida, "A 350-nm-band GaN/AlGaN multiple-quantum-well laser diode on bulk GaN," *Appl. Phys. Lett.*, vol. 107, no. 151103, 2015.
8. K. Sato, S. Yasue, K. Yamada, S. Tanaka, T. Omori, S. Ishizuka, S. Teramura, Y. Ogino, S. Iwayama, H. Miyake, M. Iwaya, T. Takeuchi, S. Kamiyama, and I. Akasaki, "Room-temperature operation of AlGaN ultraviolet-B laser diode at 298 nm on lattice-relaxed Al0.6Ga0.4N/AlN/sapphire," *Appl. Phys. Express*, vol. 13, no. 031004, 2020.
9. T. Hayashi, Y. Kawase, N. Nagata, T. Senga, S. Iwayama, M. Iwaya, T. Takeuchi, S. Kamiyama, I. Akasaki, and T. Matsumoto, "Demonstration of electron beam laser excitation in the UV range using a GaN/AlGaN multiquantum well active layer," *Sci. Rep.*, vol. 7, 2017.
10. T. Wunderer, J. Jeschke, Z. Yang, M. Teepe, M. Batres, B. Vancil, and N. Johnson, "Resonator-length dependence of electron-beam-pumped UV-A GaN-based lasers," *IEEE Photonics Technol. Lett.*, vol. 29, no. 16, pp. 1344–1347, 2017.
11. K. L. Shaklee and R. F. Leheny, "Direct determination of optical gain in semiconductor crystals," *Appl. Phys. Lett.*, vol. 18, no. 11, pp. 475–477, 1971.
12. Y. Kotsar, B. Doisneau, E. Bellet-Amalric, A. Das, E. Sarigiannidou, and E. Monroy, "Strain relaxation in GaN/AlxGa1-xN superlattices grown by plasma-assisted molecular-beam epitaxy," *J. Appl. Phys.*, vol. 110, no. 033501, 2011.
13. S. Cuesta, Y. Cure, F. Donatini, L. Denaix, E. Bellet-Amalric, C. Bougerol, V. Grenier, Q. M. Thai, G. Nogues, S. T. Purcell, L. S. Dang, and E. Monroy, "AlGaN/GaN asymmetric graded-index separate confinement heterostructures designed for electron-beam pumped UV lasers," *Opt. Express*, vol. 29, no. 9, pp. 13084–13093, 2021.
14. J. Pastrňák and L. Roskovcová, "Refraction Index Measurements on AlN Single Crystals," *Phys. Status Solidi b*, vol. 14, no. K5, 1966.
15. T. Kawashima, H. Yoshikaza, and S. Adachi, "Optical properties of hexagonal GaN," *J. Appl. Phys.*, vol. 82, no. 7, p. 3528, 1997.
16. M. G. R. Martens, "Optical gain and modal loss in AlGaN based deep UV lasers," TU Berlin, 2018.
17. W. Guo, Z. Bryan, J. Xie, R. Kirste, S. Mita, I. Bryan, L. Hussey, M. Bobea, B. Haidet, M. Gerhold, R. Collazo, and Z. Sitar, "Stimulated emission and optical gain in AlGaN heterostructures grown on bulk AlN substrates," *J. Appl. Phys.*, vol. 115, no. 103108, 2014.
18. Q. Guo, R. Kirste, S. Mita, J. Tweedie, P. Reddy, B. Moody, Y. Guan, S. Washiyama, A. Klump, Z. Sitar, and R. Collazo, "Design of AlGaN-based quantum structuresfor low threshold UVC lasers," *J. Appl. Phys.*, vol. 126, no. 223101, 2019.
19. J. F. Muth, J. H. Lee, I. K. Shmagin, R. M. Kolbas, H. C. Casey Jr, B. P. Keller, U. K. Mishra, and S. P. DenBaars, "Absorption coefficient, energy gap, exciton binding energy, and recombination lifetime of GaN obtained from transmission measurements," *Appl. Phys. Lett.*, vol. 71, no. 18, p. 2572, 1997.
20. L. Dal Negro, P. Bettotti, M. Cazzanelli, D. Pacifici, and L. Pavesi, "Applicability conditions and experimental analysis of the variable stripe length method for gain measurements," *Opt. Commun.*, vol. 229, pp. 337–348, 2004.
21. C. A. Klein, "Bandgap Dependence and Related Features of Radiation Ionization Energies in Semiconductors," *J. Appl. Phys.*, vol. 39, p. 2029, 1968.